# "BRING IT ON": EXPLAINING PERSISTENCE IN SCIENCE AT THE INTERSECTION OF IDENTITY AND EPISTEMOLOGY

LUKE D. CONLIN
*430 Wallenberg Hall, Bdg. 160*
*450 Serra Mall*
*Graduate School of Education*
*Stanford University, Stanford, CA 94303, USA*
*luke.conlin@gmail.com; 240-271-7665*

JENNIFER RICHARDS
*122M Miller Hall, College of Education, University of Washington, Seattle, WA 98195, USA*

 AYUSH GUPTA
*1320 Toll Building, Department of Physics, University of Maryland, College Park, MD 20740, USA*

ANDY ELBY
*2226N Benjamin Hall, College of Education, University of Maryland, College Park, MD 20740, USA*

**This manuscript is currently in the review process.**

## ABSTRACT

Research has documented a sharp decline in students' interest and persistence in science, starting in middle school, particularly among students from underrepresented populations. In working to address this problem, we can learn a great deal from positive examples of students getting excited about science, especially students who were previously disengaged. In this paper, we present a case study of Estevan, an 8th grade student who came into Ms. K's science class with a reputation as a potential "problem student," but left as a leader of the class, even making plans to pursue a career in science. Through analysis of interviews and classroom interactions, we show how Estevan's love of science can be partially explained by an alignment between his identity as a lover of challenges and his epistemology of science as involving the challenge of figuring things out for yourself. This alignment was possible in part because it was supported by his caring teacher, who attended to his ideas and constantly challenged him and the rest of her students to figure things out for themselves instead of just "giving them the answers."

*"When you're thinking about something that you don't understand, you have a terrible, uncomfortable feeling called confusion. It's a very difficult, unhappy business. And so most of the time you're rather unhappy, actually, with this confusion. You can't penetrate this thing…"*
—Richard Feynman, 1963

*"I mean that's something about me, like about me, I like challenges though. I mean, I like challenge myself almost all the time. I mean, last time I was challenging myself to find out what it was, I mean I'm not gonna quit…so easily"*
—Estevan, 8[th] grade, describing why he pushes through frustration to figure things out

As exemplified by Feynman, a prominent theoretical physicist, doing science requires a good deal of persistence in the face of seemingly negative emotions. Science intimately involves figuring out answers to new questions – a process that can be exhilarating, but that also involves navigating feelings like confusion, disappointment, and frustration. How scientists and, critically for this paper, science students take up these experiences and emotions plays an important role in their engagement and persistence in science.

A growing body of research suggests that young people lose interest in science over the course of their schooling (Jenkins & Nelson, 2005; Lyons, 2006; Osborne & Collins, 2001; Sjøberg & Schreiner, 2005), especially after the age of 10 (Murphy & Beggs, 2005). Science is particularly failing to engage certain populations of students, such as females and minorities (Huang, Taddese, & Walter, 2000; Mau, 2003). As a result, the scientific disciplines have suffered from a well-documented lack of gender and ethnic diversity (e.g., Medin & Bang, 2014). Regardless of whether the students go on to pursue a career in science, they are missing out on understanding what science is all about – including the joy and excitement of figuring out something new about the natural world. For these and many more reasons, an active question in the field of science education is how we might improve the engagement of young learners in science – especially students from populations who are traditionally underrepresented in science.

One way to further our understanding is to find and analyze exemplary cases of students who get turned on to science, unpacking what contributes to their shifts. This paper focuses on one such case study of Estevan, an eighth-grade student, recently emigrated from Central America, who came into the year with a reputation for potentially being a "trouble" kid but left as the leader of his science class, even making tangible plans to pursue a career in science. Through analysis of classroom interactions, as well as interviews with Estevan and with his teacher, Ms. K., we seek to understand Estevan's persistence in his pursuits in science. In Estevan's case, we argue that his engagement in science largely stems from a connection between a particular personal epistemological stance toward science, namely that science involves *the challenge of figuring things out for yourself,* and his identity as *someone who faces challenges head-on.* This alignment was supported by Estevan's interactions with Ms. K and the particular characteristics of the learning environment she set up in her classroom.

In what follows, we first review the literature on students' persistent engagement in science, focusing on research that seeks to explain persistent engagement in terms of students' identities in science. This literature suggests that students become engaged in science when the topics studied or ways of thinking connect with their cultural identities; we argue in this paper that more personalized connections may also exist and contribute to student engagement. Then, we present the case of Estevan, who initially caught our eye with his persistence in explaining how we have seasons despite mounting frustration.

The data analysis is split into three parts. In Part 1, we report on Estevan's pursuit of understanding the seasons through classroom data and field notes, and we present data from interviews with Estevan and with Ms. K that helps to explain this persistence in terms of an alignment between Estevan's identity and an epistemological stance he takes toward science. In Part 2, we explore the classroom environment that supported Estevan in arriving at this alignment, through the analysis of Estevan's classroom activity, focusing on his interactions with Ms. K and with his fellow students. We show that Ms. K and Estevan develop a caring relationship that incorporates a mutual challenging dynamic, and that Ms. K supports Estevan in recruiting this challenging dynamic in engaging in the practices of scientific argumentation and sensemaking. We conclude by discussing some implications for instruction and for research.

## EXPLAINING PERSISTENT ENGAGEMENT IN SCIENCE

A central goal in science education is to encourage all students' continued interest and engagement in science, for many reasons, from the recruitment of more and better scientists to the fostering of an informed critically thinking citizenry that considers (scientific) evidence when voting on issues. But the choice to pursue science as a career is not presently equally accessible to all students, as evidenced by the well-documented underrepresentation of women and minorities in science majors and careers (Blickenstaff, 2005; Cole & Espinoza, 2008; Huang et al., 2000; Mau, 2003; NCES, 2000). As such, a critical goal in science education is to encourage underrepresented students' continued interest and engagement in science.

There is a growing body of research on students' interest, engagement, and persistence in science, some of which focuses on students who are traditionally underrepresented in science. Some of this work has emphasized helping underrepresented students pursue scientific topics that connect to their cultural backgrounds (Hammond, 2001) or other aspects of their lived experiences and identities (Calabrese Barton, 1998; Basu & Calabrese Barton, 2007; Seiler, 2001). Other work has emphasized the importance of making epistemological connections to promote minority students' engagement in science by incorporating culturally-relevant ways of thinking and interacting (Bang & Medin, 2010; Hudicourt-Barnes, 2003; Warren, Ballenger, Ogonowski, Rosebery, & Hudicourt-Barnes, 2001). In most of this previous work, attention has focused on connecting classroom experiences (conceptual or epistemological) to students' ethnic or cultural identities and backgrounds.

### Engaging Students In Science Through Identity-Based Connections

A growing body of research across multiple methodologies seeks to explain students' engagement and persistence in science in terms of their identities (Calabrese Barton & Yang, 2000; Basu, 2008; Brotman & Moore, 2008; Carlone & Johnson, 2007; Chinn, 2002; Cleaves, 2005; Dabney, Chakraverty, & Tai, 2013; Gillenbrand, Robinson, Brawn, & Osborn, 1999; Hazari, Potvin, Tai, & Almarode, 2010; Hughes, 2001; Lee, 1998, 2002; Morgan, Isaac, & Sansone, 2001; Stokking, 2000). For instance, O'Brien,

Martinez-Pons, and Kopala (1999) conducted statistical path analysis of four-hundred and fifteen 11[th]-grade parochial school students' self-efficacy, gender, mathematics self-efficacy, academic achievement, reported ethnicity, and socioeconomic status. The resulting path model held that ethnic identity influenced mathematics self-efficacy; mathematics self-efficacy was then predictive of career interest, which was in turn predictive of persistence in the pursuit of scientific careers (Schoon, 2001). Here, the relevant aspect of identity was the individual's ethnic origin, taken as a static characteristic.

Other studies have adopted ethnographic methods to document and explain heightened engagement in science and math through the dynamic construction of identities that bridge students' senses of self with their disciplinary activities. In a study of urban youth ages 10 to 14 engaged in a year-long voluntary green energy program, Calabrese Barton and Tan (2010) documented how the students' blending of personal and cultural identity with science identity into a "community science expert" gave the students "a platform in which to engage in scientific ideas and discourses while also offering students the freedom to work and be in their community in ways that mattered to them" (p. 221). Nasir and Hand (2008) used close ethnographic observations across different settings to explain two African American high school students' differential engagement in math class. The analysis centered on how a mathematics classroom differentially afforded two students' formation of identities that were linked with the practice of mathematics in ways that gave them access to the domain, integral roles within the domain, and opportunities for self-expression. In Calabrese Barton's and Nasir's studies, identity is treated as being more fluid and constructed, with multiple interacting dimensions (including a cultural dimension and a disciplinary dimension).

The studies above point to the relevance of learners' identities to disciplinary engagement and persistence in science, but they also highlight differences in how identity is framed and treated as a theoretical construct. Others have noted that identity is often used as a construct without being clearly defined (Brown, 2004; Nasir & Hand, 2008; Stets & Burke, 2003), and the treatments above range from a sense of identity as a set of trait-like aspects of the individual (e.g., O'Brien, Martinez-Pons, & Kopala, 1999) to a sense of self developed in interactions and relationships with others (e.g., Nasir & Hand, 2008) to a dynamic positioning within particular activity structures (e.g., Esmonde, 2009). It will be worthwhile to clarify what we mean by "identity" and how we look for it empirically. In the Methods section, we outline a broad perspective on identity that focuses on the common ground amongst the various treatments in the literature, and that affords empirical examination through a variety of methods and methodologies.

Additionally, our study addresses a gap in the extant literature reviewed. The linking of identities to practices is an important contribution on which we build. However, in both of these studies the notion of what counts as disciplinary practices needs to be filled in. We will show that Estevan's identity gets linked to the practices of making sense of physical phenomena, and to persisting through the struggle to make sense of things for himself. One challenge in creating equity in science education is that minority students have limited access to schooling that is supportive of critical thinking and high quality science instruction that is authentic to disciplinary practices. As Calabrese Barton & Tan (2010) note:

Many youth from low-income communities do not have direct access to traditional networks of resources, such as experts in the field or materials, and when they do have access they are often positioned as recipients of the expertise rather than participants in the use and further construction of expertise (Oakes, 1990, 2000) (p. 190).

It is thus critical that when addressing the need to promote and retain minority students' interest in science, emphasis and attention is on providing students with opportunities to engage in authentic disciplinary practices of science – engaging in generating causal explanations for natural phenomenon (Hammer & van Zee, 2006), empirical reasoning (Warren, Ballenger, Ogonowski, Rosebery, & Hudicourt-Barnes, 2001), and design of experiments (Etkina).

### Engaging Students In Science Through Culturally-Based Epistemological Connections

Another approach to promoting minority students' engagement in science is incorporating culturally-relevant ways of thinking and interacting in the science classroom (Bang & Medin, 2010; Hudicourt-Barnes, 2003; Warren et al., 2001). Bang and Medin (2010) discussed curricula organized around an epistemological stance that living things are all related; this epistemological orientation is rooted in practices and beliefs of the Indigenous communities for whom the curriculum was being designed. Students come into the science classroom with many ideas about what science is, how it works, and how to learn it. For instance, a student may view science as a collection of disconnected formulas to be memorized, or as a set of connected concepts to understand (Hammer, 1994), and often the same student will switch back and forth between these ways of viewing science. We refer to an individual's views towards knowledge and learning as their "personal epistemology" (Hofer & Pintrich, 1997). A growing body of research has demonstrated that these personal epistemologies ideas can influence how students approach thinking and learning in science class (Hofer & Pintrich, 1997; Sandoval, 2005).

Our case of Estevan adds to the extant literature by expanding beyond a focus on epistemological stances associated with students' ethnic identities to consider more individualized connections between aspects of students' identities and their learning in and stance toward science. For Estevan, his interest in science is sustained not through topical connections, or connections to cultural epistemology, but through a coherence between his identity as someone who loves challenges and his sense of science as a pursuit of figuring stuff out for oneself. It is this personalized connection that is nurtured through his experiences in his eighth-grade science classroom. By presenting the case of Estevan, we want to contribute to the literature by 1) emphasizing the need for greater attention to epistemology in providing students with authentic disciplinary practices in science, 2), demonstrating how epistemic practices of science such as inquiry and argumentation can be a hook for students, including some who might otherwise be turned away from science and 3) arguing that in some cases pathways to authentic disciplinary engagement in science might be more individualized than previously considered in the literature.

## METHODS

In this section, we provide relevant contextual information about the settings from which the data came. We also describe the data sources we drew on in understanding Estevan's persistence, as well as our analytical approaches.

### Context

The data for this study come from a professional development project aimed at helping fourth through eighth grade teachers promote inquiry teaching and learning in their science classrooms. Specifically, inquiry is promoted as primarily a matter of creating causal, clear, and coherent explanations of scientific phenomena. As part of the project, teachers attend a two-week summer workshop in which they engage in their own minimally-guided inquiry, watch classroom video of students discussing scientific phenomena, and collaborate on other issues related to inquiry teaching and learning in the classroom (i.e., assessment, lesson planning, etc.). During the school year, teachers work one-on-one with members of the research team to facilitate scientific inquiry in their classrooms and attend bimonthly small group meetings with other teachers and members of the research team.

At the time of the study, Ms. K was an experienced teacher in her first year with the project. She taught seventh and eighth grade science at a Title I middle school with a predominantly Hispanic and African American student population. Jen (the second author) regularly visited Ms. K's classes as part of the project and noticed Estevan's persistence in working through and trying to understand a puzzling scientific phenomenon; a full account of this focal episode is provided in Part 1. The research team sought to understand this episode as an exemplar case of student persistence in sense-making, and ultimately Estevan as an example of a student who got turned on to science as a possible career. Additionally, Estevan's status as a member of an underrepresented population in science and as an English language learner provided us with an opportunity to contribute to the body of knowledge on underrepresented students' engagement in science.

### Data Sources

The data for this case study consist of several videotaped classes of Ms. K's from the 2011-2012 school year, as well as videotaped interviews with Estevan and Ms. K about the focal episode described in Part 1. The focal episode occurred partway through the school year in January. Building on Jen's field notes and video of the focal episode, she conducted semi-structured interviews with both Estevan and Ms. K to glean their perspectives on what happened in class. With Ms. K, the interview also included watching video of the episode to stimulate her recall of important events (Lyle, 2003). Preliminary analysis of these data sources (see Richards, Conlin, Gupta, & Elby, 2013) highlighted the importance of the constructs of identity and personal epistemology in Estevan's case. Here, we expand on this initial work and consider Estevan's participation in other class sessions, both to explore the roles of identity and epistemology in the classroom context and to understand more about the dynamics that may have contributed to his engagement in sense-making.

**Discussion on Methodology**

As indicated, our analysis draws on the constructs of identity and personal epistemology. These constructs, however, have been described and investigated in a variety of ways (at times, in conflict with each other) in the science education research literature. So, first we present a brief outline of some of the methods used for investigating identity and epistemology, before outlining our methodological inclinations and analytical approach.

*Investigating Identity*

Identity has been an important construct across many disciplines, from developmental psychology (Guardo & Bohan, 1971) to sociology & social psychology (Burke, 2006; Stets & Burke, 2003), sociolinguistics (Bucholtz & Hall, 2005), and education (Gee, 2001). Within education, there is a growing body of research on students' identities and their role in the learning of science (Aschbacher, Li, & Roth, 2009; Calabrese Barton & Tan, 2010; Brickhouse, Lowery, & Schultz, 2000; Brown, 2004; Carlone & Johnson, 2007; Hazari, Sonnert, Sadler, & Shanahan, 2010; Lemke, 2000; Nasir & Hand, 2008; Roth & Tobin, 2007; Wenger, 1998; Zimmerman, 2012).

However, literature within and across these disciplines differs in its characterization of the construct of identity. Some treat identity as a conceptual entity, i.e., identity as the concept a person has of their own individuality, including defining characteristics, social roles, abilities, and worth (e.g., Stets & Burke, 2003). Others maintain that identity is only definable in terms of social structures and relations (e.g., Burke, 2006) or that it emerges through linguistic interaction (e.g., Bucholtz & Hall, 2005). Research within education has drawn from these different characterizations of identity, often incorporating several seemingly distinct elements within the same account of identity. For example, Gee (2000) distinguishes between a "core identity" that holds more uniformly for the self and others and a shifting, multi-layered identity that gets produced through interactions with others. These two elements, which we refer to as *sense of self* and *enacted identity* respectively, can be found across much of the literature on identity in science education and beyond. We briefly review each element in turn below before describing how we treat identity in this study.

*Sense of self*

Sense of self is an aspect of identity that cuts across most approaches to characterizing and empirically tracking identity. Some of these approaches tend to *equate* sense of self with identity (e.g., Hazari, Sonnert, et al., 2010), while others incorporate sense of self as one of the important aspects of identity (Bucholtz & Hall, 2005; Gee, 2001). Generally speaking, sense of self corresponds to a person's experience or conception of themselves as having certain characteristics (Guardo & Bohan, 1971). It is one's self-recognition of being a certain "kind of person" in a given context (Gee, 2001; Hazari et al., 2010). This sense of self could encompass sub-constructs that are conceptual (self-concept) and/or affective (self-esteem) in nature (Burke, Owens, Serpe, & Thoits, 2003). In science, researchers have focused on whether students identify as a "science person" (Carlone, 2004) or with a particular subdiscipline, such as Hazari et al.'s (2010) work on physics identity. Interviews and survey instruments are commonly used to access this aspect of identity (e.g., Aschbacher et al., 2009; Hazari, Sonnert, et al., 2010), as self-reports of being a certain kind of person or of having certain characteristics is taken as good evidence of one's sense of self.

### Enacted identity

A common problem, though, with self-reports of identity is that people may behave in ways that are quite distinct from, or even counter to, how they view themselves. This has led some researchers to focus on what we are calling *enacted identity* – how identities are established through words and actions (Bucholtz & Hall, 2005; Gee, 2010; Nasir & Hand, 2008). Work in this vein primarily focuses on the discourse and interactions through which identities are constructed, as people author themselves and/or are authored by others (Holland, Lachiotte Jr., Skinner, & Cain, 1998). For example, Gee (2000) recounted a scenario from a second grade classroom in which one girl used words and actions that served as "bids" to get herself recognized as an enthusiastic learner. First she made a bid by proactively correcting errors in a spelling task they hadn't gotten to yet, and then by bouncing in her chair to enthusiastically volunteer for an activity in which the teacher chose students to try to predict the story of a book based on pictures. Her bids were unsuccessful in each case, as the teacher responded by scolding her for jumping ahead in the spelling activity, then told her to calm down as she tried to volunteer—discourse moves that positioned the learner as the sort of student who needs to be managed by the teacher according to curricular procedures. Enacted identity is best studied through close examination of interactional moments like these.

### Investigating Personal Epistemology

Similar to the discussion of identity above, researchers have treated learners' epistemologies or stances toward science in different ways. Some have analyzed students' epistemologies in terms of *beliefs* that are robust and hard to change, accessed through self-reporting in interviews and on questionnaires and survey instruments (Hofer & Pintrich, 1997; Perry, 1970; Pomeroy, 1993; Schommer, 1993). In their influential work on students' personal epistemologies, Hofer and Pintrich (1997) discussed how students can take particular stances (e.g., absolutist vs. relativistic) with respect to aspects of scientific knowledge, such as its certainty and its source. They did not explicitly define what they meant by "stance," but used the term to characterize large-scale categories of beliefs about knowledge.

Other research has focused on more fine-grained aspects of epistemology – more contextually-sensitive, less static, and inferred through close analysis of students' reasoning and behavior in context. For instance, Hammer and Elby (2002) laid out a framework of students' epistemological resources for thinking about scientific knowledge, including resources for thinking about the nature of knowledge, activities that produce knowledge, and forms such knowledge can take. Within this perspective, stances can reside at much smaller grain-sizes than those implied by Hofer and Pintrich (1997), such as at the level of *doubting* a particular claim in the moment. diSessa, Elby, and Hammer (2002) have provided a way of bridging these larger and smaller scale perspectives. They explained a physics learner's behavior and reasoning in a clinical interview by inferring several aspects of her epistemological stance. Based on patterns in the interviewee's behaviors during the interview, such as a tendency to hedge her claims or to forward contradictory claims, the authors made a claim about her larger-scale epistemological stance as including a tendency to view learning physics as sense making. This process reflects a bottom-up approach where smaller grain-sized stances reflect patterns that give rise to larger grain-sized stances.

### *Our Theoretical Commitments*

Our approach to investigating identity and personal epistemology in this study generally aligns with the bottom-up approach proffered by diSessa, Elby, and Hammer (2002). We begin with an assumption that one's identities and epistemological stances are fluid and context-dependent – that one may author oneself or be authored in various ways in different moments, and one may have various ways of thinking about scientific knowledge and learning under different circumstances. If accumulations or patterns of identifications or epistemological stances are evident in data over time and across contexts, such patterns may constitute a sense of self or more general epistemological belief[1]. What is important to note here is that we are open to and in fact regularly expect variability; any consistencies we describe are empirical results, rather than presumed stabilities.

## Analytical Flow

There were three general phases to our analytical process, which we discuss in turn. For each phase, we describe the aim of the phase (what we were trying to understand) and the techniques we used in working with the data.

### *Part 1*

Part 1 includes a largely descriptive analysis of the focal classroom episode in which Estevan persisted in making sense of a puzzling scientific phenomenon. In this phase of analysis, we drew on Jen's field notes, classroom video, and interviews with Estevan and Ms. K to flesh out what Estevan was trying to understand and how he was going about doing so. We triangulated across these data sources to create a narrative of the focal event, establishing that Estevan's frustration was centered on reconciling 1) his idea that the Earth's tilt causes the seasons with 2) a textbook diagram showing that the Northern and Southern hemispheres receive the same amount of sunlight in the Spring and Fall. We also noted the nature of Estevan's activities during class, which alternated between working with the textbook and a globe in relative solitude and asking Ms. K questions (only to be given more questions in return).

### *Part 2*

In Part 2 of our analysis, we focused on understanding *why* Estevan persisted in class the way he did. Here, we worked primarily with Estevan's interview video, with an eye toward how he experienced the focal episode and what stood out to him. Several themes emerged in this phase of analysis (see Richards, Conlin, Gupta, & Elby, 2013), most notably Estevan's positioning of himself in the interview as someone who loves challenges and his description of science and Ms. K's class as places where you have to figure things out for yourself. In the interview context, Estevan's positioning was coordinated across multiple modalities (Jordan & Henderson, 2012; Stivers & Sidnell, 2005) as he made direct statements about himself and (re)enacted a challenging interplay between himself and Ms. K in which he adopted a cocky tone and body language.

---

[1] Furthermore, more "stable" senses of self or epistemological beliefs may feed back into in-the-moment dynamics; such forces are likely in constant interplay.

These themes cued us into the constructs of identity and personal epistemology and a specific entanglement between the two in Estevan's case, as he described figuring things out in science *as* a challenge. However, we recognized that the particular stances Estevan adopted may be limited to the interview context and interpersonal dynamics between Estevan and Jen, though there was some evidence from his storytelling about other contexts that they extended beyond the proximal setting.

The interview with Estevan also provided clues about features of the classroom environment and Estevan's interactions with Ms. K and other students that may have contributed to his sustained pursuit. The next phase of analysis builds from these insights, seeking to understand what it was about Ms. K's class that sparked his engagement.

### Part 3

In the final analytical phase, we explored other classroom video we had from Ms. K's class over the course of the 2011-2012 school year for several purposes. First, having hypothesized a coherence between Estevan's love of challenges and an epistemological stance of figuring things out in science in the interview context, we aimed to see if there was confirming or disconfirming evidence of this coherence in the classroom. We focused on Estevan's positioning as described above – through statements he made about himself and others (though these were less plentiful in the flow of classroom activity) and ways in which he interacted with Ms. K and his peers. For instance, if Estevan said something that put someone else's face on the line (Goffman & Best, 2005) in a serious tone of voice, followed by laughter, we interpreted that he was playfully challenging the person. We then considered what Estevan was challenging or being challenged with respect to. We inferred epistemology based on how Estevan acted during class activities – for instance, if he was actively engaged in constructing an explanation for a scientific phenomenon or trying to solve a problem, such activities would be consistent with an epistemological stance toward figuring things out. On the other hand, if Estevan asked for someone to tell him the answer, that would not be consistent with figuring things out for yourself. While we sought to understand examples from the classroom in which either dynamic was at play – challenging or figuring things out – we focused most intently on examples in which both dynamics seemed to be at play and the interrelations between the two.

Second, we wanted to understand more about the classroom and interactional features that may have supported Estevan in engaging and persisting in sense-making. To this end, we focused primarily on classroom interactions that occurred between Estevan and Ms. K prior to and during the focal episode, positing aspects of their relationship and the broader classroom culture that may have contributed to his engagement and considering alignment between the classroom data and what Estevan noted in the interview.

## DATA & DISCUSSION – PART 1

We begin by situating and describing the focal episode that originally caught our attention in detail. In the class prior to the focal episode, Ms. K facilitated a whole-class discussion by starting with a surprising fact: The Earth is closer to the Sun during the Winter[2]. Her question for them was, "How is this possible?" She told them to start with what they know about the seasons and the Earth's relationship to the Sun or within the solar system. The ensuing discussion focused mainly on what causes the seasons, with two main ideas emerging that were in competition during the debate. The main disagreement was over whether it the "tilt" that causes the seasons (as Estevan argued), or the "rotation" (as another student, Reggie, argued). According to Estevan, it is the tilt that makes one side closer or not, but for Reggie it is the Earth's rotation that determines which parts of the Earth are closer to the Sun at any given time. Previous research has demonstrated that making sense of the seasons is a difficult matter for middle school students and even extensively-educated adults (for a brief review, see Sherin, Krakowski, & Lee, 2012).

Toward the end of class, Ms. K instructed students to split into groups and create models of how interactions between the Earth and Sun result in the seasons (including how the Earth can be closer to the Sun in the winter). Students were able to work with a variety of resources, including blow-up globes and their textbooks, and model construction continued into the next class meeting.

**Estevan's Conundrum**

As students were working on their models on the second day, Estevan repeatedly approached Ms. K with ideas and questions. At one point, he brought a diagram from the textbook to Ms. K that depicted the Earth at various places in its orbit around the Sun: December, March, June, and September. Jen's field notes depict this particular conversation between Estevan and Ms. K as follows:

> I got the impression that Estevan was trying to make sense of the representation, and he seemed to be comfortable with the idea that in our winter, the Northern Hemisphere is tilted away from the Sun, and in our summer, the Northern Hemisphere is tilted toward the Sun. He asked if that was right, and Ms. K wouldn't tell him whether that was right or wrong, but asked if it made sense to him... [and told him that] he's doing a fantastic job of figuring this out, and he should trust whether it makes sense to him or not. He said it did, but that something was missing. He wasn't sure what, but something was missing. He had his head in his hands and was kind of staring at the book at this point. He asked if Ms. K could give him a clue, and she asked him what was going on in March. He said that was more like a riddle... [Field notes, January 19, 2012].

In this exchange, Estevan attempted to get Ms. K to give him information, but she refocused him on his own sense-making. Estevan acknowledged that the Earth's tilt in winter and summer made sense to him, but "something was missing." He seemed somewhat frustrated, putting his head in his hands and continuing to ask for information. Again, Ms. K did not provide information, but rather asked him about the spring.

---

[2] It's unclear whether she specified that this is for the Northern Hemisphere, but they get to that in the discussion.

Throughout much of the 90-minute class period, Estevan cycled between discussions like this with Ms. K and retreating to his seat to think further. At his seat, he alternated between referring to pictures of the seasons in the textbook and maneuvering a blow-up globe in different configurations. At times, he engaged in conversation with other students, but he primarily spent this time in seclusion. Ms. K noted this behavior in her interview:

1. Ms. K: You know when he would go back over into his little solitude, and he would-
2. Jen: With the book.
3. Ms. K: Think it through- mm-hmm, with the book.
4. Jen: I remember him just staring at the book, at points.
5. Ms. K: Mm-hmm, and he would have the globe and would look at- and he would manipulate it. And he would come back, and he would say, "Okay, it's this." [Ms. K interview, January 24, 2012]

According to Ms. K, Estevan was specifically trying to reconcile how the Earth could be tilted in the fall and spring when both hemispheres were receiving equal amounts of sunlight:

Ms. K: He, he couldn't figure out for – the spring and the fall, how the tilt- he didn't know if, if, say we're the Northern Hemisphere, in the fall well is it here ((holds hand at an angle)), here ((holds hand straight up and down)). He wanted to think it was straight up, but he knew the axis was in there, and he couldn't figure out exactly how the axis was in, in relation to the Sun [Ms. K interview, January 24, 2012].

Estevan initially made the case that the Earth was straight up and down in the fall and spring. Ms. K recalled his explanation as having to do with the amount of Sun each Hemisphere gets: "I remember him saying to me at one point, 'But it gets the same amount of Sun, so the Sun would hit like this (holds blow-up globe level), and it would be straight. So it's gotta be straight.'" Yet he was stumped by the textbook picture, which depicted an axis that was always tilted. Ms. K indicated that he kept coming back to her, and "his focus was March and September – over and over and over." He asked her for clues and questions, and she responded by asking him questions that pressed on his current understanding (e.g., "what was going on in March" according to Jen's field notes, or "look back at the axis, see where that fits in" according to her interview).

What was most notable to Jen about Estevan's behavior throughout the class period was his persistence in the face of evident frustration. Jen brought this observation up in the interview with Ms. K:

6. Jen: I guess I felt like he was, like got frustrated? But productively so.
7. Ms. K: Oh, he was-
8. Jen: Was that your sense?
9. Ms. K: He wouldn't give up. He would not give up. He was like, "I'm gonna understand this," and then he would get like ((hunches a bit and looks off into space)), but then he would go right ((snaps)) back ((snaps)) to it ((snaps)) [Ms. K interview, January 24, 2012].

In this exchange, Ms. K portrayed Estevan's vacillation between frustration (and perhaps a bit of dejection, folding into himself) and persistence. She indicated that the core of the issue was conceptual in nature, about what Estevan felt he did or did not understand. Later, Ms. K also suggested that Estevan may have gotten frustrated with her responses, but he never gave up:

> Ms. K: I never told him either you're right or wrong. Or I'd say, "seems like you're on the right track." And – I think a little bit of that might have frustrated him too? But he would always come back for more. He would always go back, and – think it through [Ms. K interview, January 24, 2012].

**Reaching a Resolution**

Ms. K reported that Estevan came to an important realization toward the end of the class period: "He started to process that it wasn't tilting forward or backward [toward or away from the Sun], it was tilting from side to side." This understanding was also reflected when Ms. K asked Estevan to present his group's model to the class. Estevan sequentially demonstrated how the Earth was tilted relative to the Sun in each season, and he described Fall as follows: "That's Fall, cause here like, um, March, we still get the same amount of, um, light, sunlight, and we're not actually tilted toward the Sun." As he was talking, he initially tilted the blow-up globe toward the "Sun" (a yellow balloon), but then changed the tilt so that it was not toward or away from the Sun, but tangential to the Sun. This coheres with Ms. K's sense that he came to see the Earth as tilting from side to side relative to the Sun in the Fall and Spring, and incorporates his earlier idea that the Northern and Southern Hemispheres would both receive the same amount of Sun in these seasons.

Estevan was then asked to repeat his group's model, and Ms. K probed in more depth about the tilt in the Fall:

10. Estevan: In June, the North side is actually in Summer and, um, the Earth is actually tilted toward the Sun. The North side is getting a lot of Sun. The other, um, part, the South, is not actually close to the Sun, which means it's winter. Um, keep rotating and rotating until we get to, um, September, that means that we are in Fall. The Earth is not actually tilted toward the Sun, it's actually tilted the other way.
11. Ms. K: Oh, it's tilted away from the Sun.
12. Estevan: It's not actually away from the Sun, it's kind of like not really close to the Sun because both, both like North and South, are actually getting the same amount of energy from the Sun.
13. Ms. K: So show me how that would be in relation to the Sun.
14. Estevan: Um, that actually would be like tilted this way ((tilts globe sideways)), like they're both getting the same amount ((gestures toward Northern and Southern hemispheres)). It's not actually tilted that way ((tilts globe toward balloon)) or the other way ((tilts globe away from balloon)) [Classroom video, January 19, 2012].

Again, Estevan tilted the globe sideways relative to the balloon and indicated that the Northern and Southern Hemispheres were "getting the same amount of energy from the Sun" (line 12). In total, the group's model (presented only partially here) provided a correct, functional model of how the Earth and Sun were positioned relative to each other to create the seasons and attempted to account for how the Earth is closer to the Sun in the Winter.

## Part 1 – Summary & Discussion

The focal episode described above depicts the struggles of a middle school student, Estevan, who persisted in the face of frustration for the better part of a 90-minute class period. His frustration was primarily conceptual in nature, as he attempted to make sense of how the Earth could be tilted in the Fall and Spring when both Hemispheres receive equal amounts of Sun. His frustration was also partly interactional in nature, as his teacher, Ms. K, did not tell him anything definitive but rather asked him questions that spurred continued thought. Despite these frustrations, Estevan persisted in sense-making until he reached a satisfying reconciliation and, ultimately, a mostly accurate model of how the seasons work.

We would not presume that a student's epistemological stance would be stable across contexts or from moment-to-moment. Indeed, Estevan does not exhibit stable epistemological framings. He asks for answers at times, and dives eagerly into figuring out mechanisms on his own at other times. His "stance" of figuring things out is interactively constructed even though his persistence and continued attempts to make sense definitely seem to indicate a reliable pattern.

We next turn to why Estevan persisted in the way that he did throughout the class. An extended interview with Estevan uncovered several important themes related to his interest and persistence in the seasons discussion and science more generally. An interview with Ms. K corroborated these themes.

## DATA & DISCUSSION – PART 2

### Understanding Estevan's Persistence

In part 2, we report on an interview conducted by Jen with Estevan a few days after the focal episode. The interview revealed that Estevan's persistence was stabilized by the intersection of his identification of someone who loves a challenge and his epistemological stance towards science as involving the challenge of figuring things out for yourself. It also revealed that this way of viewing science had roots in his interactions with his father, and has been rekindled by Ms. K's approach to teaching.

### *For Estevan, science involves figuring things out for yourself.*

At the very beginning of the interview, Jen said "it was pretty cool to watch the other day, what was going on," referring to the events of the focal episode. In response, Estevan emphasized that part of what he liked about Ms. K's class was that he was expected to think for himself:

> I knew something was missing though. Because the explanation didn't kind of make sense. Until- you saw she didn't give me no answer, though. She just stood there and look at me like you gotta make it up, you gotta think about it. I mean, I like that about her, though. I love all that about her because she just make us think about everything. She don't give us the answer. And that's what make it fun.

Here, Estevan reiterated his sense from class that "something was missing" because what he was grappling with did not "make sense." He then described how Ms. K did not give him the answer; rather, she expected him to "think about it," which he qualified as "fun." Later in the interview, Estevan associated *figuring things out for yourself* not just with Ms. K, but with science: "She don't give us no answers, she just, 'Figure it out for yourself. And then you come to me. Then I'll correct you if you're wrong.' *I like that a lot about science though.*" (Emphasis added.) He later contrasted Ms. K's class with previous science classes in which the teachers "didn't let us do what we wanted to do with the science, you know, find out."

In these and other statements throughout the interview, Estevan demonstrated a personal epistemological stance that *science is something that should make sense, something that you can figure out.* For Estevan, this experience of science was also affectively charged. He found it "fun" when Ms. K would not give him the answer, and he described the moment when "you just found all the pieces, and you put it all together, and you're actually right" as "the best feeling you can get." Notice too that his complaint about former teachers, that they didn't "let us do what we wanted," referred not to a specific science topic but to "find[ing] out," i.e., figuring things out.

We now explore how Estevan's enjoyment of figuring things out in science connects to his more general love of challenges, which he portrayed as critical to his sense of self.

### Estevan identifies as someone who likes to face challenges.

During the interview, Jen probed why Estevan kept trying to figure out the seasons in class, despite his demonstrated frustration:

15. Jen: I've been in positions I think where I've been trying to figure something out, and I knew something was missing, and I'm trying to figure something out, and I've gotten so frustrated ((Estevan laughs)) that I actually haven't been able to, like, keep going with it? But you kept going with it, so-

16. Estevan: I mean, yeah, I mean that's something about me, like about me, I like challenges though… I mean, for me, that's, that's the thing that actually most I love. Challenges, I mean, every single day challenge that I can get- that's why I tell you ((smiles)), I mean, I, last time we were doing everything, and she was asking me questions, and I challenged her, like, ask some harder questions so I can, you know- I like challenges, that's the thing about me. I mean, I don't quit so easily though. I won't, until I find something.

17. Jen: Yeah. No, that totally makes sense- so you were actually asking her for harder questions?

18. Estevan: Yeah, I came, because she like, she told me, give me like a little review or explanation, and then she started asking questions, right? And I'm like, I answer all her questions then I'm like, ((leans back slightly and cocks head to side, see fig)) you got another question, like <u>harder</u> than that? ((both laugh)) She actually gave me a harder one, though, she

> actually told me why we're closer to the sun. That's when I'm like oh, okay, ((cocks head to side)) I'm about to figure this one out too.

The exchange above brings out the sense that tackling challenges is something that is central to Estevan's *sense of self*. His repeated references to himself, using phrases like "that's something about me" and "for me" (line 16), indicate that his love of challenges is something specific about *him,* something that he thinks sets him apart. Estevan is not simply reporting on his sense of self. He is likely presenting a certain version of himself, partly (in 16) to contrast himself with Jen. So, we are not getting a "read-out" of his sense of self, but rather a contextualized performance of this sense. But the context is talking to an adult who he may associate with Ms. K and with school science; so, to the extent his performance is tied to context, it's tied to the appropriate context for the purpose of our argument.

As Estevan talked about challenging Ms. K to ask him harder questions, and as he recapped their interactions, his physical movement of leaning back slightly in his chair and cocking his head to the side demonstrated a sense of bravado as he asked if Ms. K had another question "<u>hard</u>er than that?" (line 18) Additionally, in the interview with Ms. K, her body language and intonation were similar as she described Estevan coming back for more – "he would say, ((juts chin up)) <u>ask</u> me another question." This suggests that Estevan was conveying a challenging stance not only upon recollection in the interview but also in his actual interactions with Ms. K during class, illustrating alignment between his sense of self and his enacted identity.

There is also evidence that for Estevan, figuring things out in science *is* a kind of challenge. Note how when Ms. K provided Estevan with a harder question (likely about being closer to the sun in the winter), his response again included a bit of bravado as he stated, "I'm about to figure this one out too" (line 34). He took her question as a challenge that he could face head-on.

Later in the interview, Estevan explicitly made the link between figuring things out in science and facing challenges:

> How is just the earth actually standing without no, um, stuff behind it, like, how is that magnetical field created about it, you know? If you actually got some questions and you need an answer, you need to put your challenge to it, get some kind of challenge because, that's for me, though, I get my challenge to find out how does that happen, learn how, why does it do that, how can it do that, what caused to do that?

Here, Estevan identified the kinds of scientific questions he wonders about, which are mostly about "how" or "why" something happens, "what caused" it to happen. He also discussed seeking answers to these sorts of questions as one way in which he "get[s] [his] challenge" – again, reiterating "that's for me, though." For Estevan, if you need an answer to a question, "you need to put your challenge to it." Thus, in this passage and throughout the interview, Estevan linked figuring things out in science to facing challenges, which he enjoys and at which he persists.

### *This kind of engagement reflects a change for Estevan.*

Coming into Ms. K's class, Estevan seemed to have a lot stacked against his persistent engagement in science. In an interview, Ms. K described Estevan as a "hard core" student who was generally "in trouble quite a bit." At the time she taught him, he had been in the United States for four years, and he was a "definite English language learner." And as Estevan tells it, he did not always feel this way about science. Although he enjoyed learning about science as a young child with his dad, he had limited interest in his previous science classes and in Ms. K's class at the beginning of the school year.

Reflecting on her interactions with Estevan shortly after the seasons discussion, though, Ms. K noted that she saw "a major difference from day one till now. Major difference." Conversations she had with Estevan also indicated a transition on his part:

> He keeps telling me that, um, he never did think science could be fun. That it wasn't- he really didn't like it, he didn't get it. He said then- that's what he told me today, he said, "Now I get it. I can understand things better."

Moreover, this transition on Estevan's part seems to have been *specific* to his science class. As Ms. K was describing the changes she saw in Estevan, Jen asked whether other teachers saw similar changes:

> 19. Jen: Is it across the board, is it more in here?
> 20. Ms. K: No, it's in here.

Though this kind of engagement is a change for Estevan, it is not entirely new. Through his interview, we learned that he has resources for thinking about science as involving the challenge of figuring things out for yourself, developed at an early age in interaction with his father. For instance, at one point Jen probed about how he asked Ms. K for harder questions, and Estevan spontaneously brought up nature walks he used to take with his father:

> I kept, I mean, I sometimes, if I'm, like, you know, she keep asking me stuff, I make her kind of push me a little bit more (?), you know what I'm saying? Like a good challenge. I mean, I love good challenges though. The thing I love about, I mean, science, for me, before it wasn't like that. I used to love science when I was a little kid though. I mean, I used to, like, me and my dad, we used to go- sometimes we used to go to this lake, you know, experiment with new plants or whatever. He used to taught me everything about science.

Later in the interview, Jen returned to what Estevan said about how he used to love science as a little kid:

> 21. Jen: So you were saying that you liked it more sort of as a kid, right, sort of like in interaction with your dad and different things?
> 22. Estevan: Yeah, I actually, what happened is that I liked it when I was a little kid, since I was a little kid, I liked finding new rocks and trying to see what was like, why were they different and stuff like that. I used to do that all the time because I used to find all kind of rocks, like, some of it were like kind of orangey and stuff, and I used to ask my dad because I didn't even know what kind of rock they were, like, I used to ask my dad what's wrong

> with the rock, why this happen on the color, and he like **you need to find out yourself**, and I started looking at them and stuff...(emphasis added)

Notable in Estevan's statement is how his father responded to his questions about the rocks – Estevan said his father told him "you need to find out yourself" (line 38), similar to the interactions Estevan reports in Ms. K's class. Estevan then explicitly connected the feeling he used to have toward science with the feeling he now has in Ms. K's class:

> But then when I got to some, what actually make me kind of forget about science is that the excitement that I had as a kid that I could do stuff with it, when we got to the classroom, some teachers actually didn't let us do what we wanted to do with the science, you know, find out. It wasn't like her because last year, like I said, science wasn't- since elementary school, science was kind of vanishing and vanishing until I got to her classroom. That's where my feeling about science came back again...

Throughout the interview, Estevan referred to Ms. K as having "brought back" his feeling toward science, saying repeatedly how he cannot thank her enough for doing that. And this feeling of science was connected to finding out for himself, which what he wanted to do but some teachers he felt did not allow for that. He contrasts Ms. K with previous teachers, who he said "didn't let us do what we wanted to do with the science, you know, find out."

Later in the interview, Estevan talked about pursuing a career in science, listing specific plans he is making: making the honor roll, enrolling in an IB (International Baccalaureate) program, getting a scholarship, and getting into college. He then reflected on how his pursuit was bolstered by having recovered this feeling toward science:

> Because I mean science, for me, it'll stay with me until I die. Like I said, science for me, it's gonna stay with me, I mean, I'm not gonna give up on science though. They already took that feeling away when I was in elementary school, and she brought it back in though, I mean, like something that you found, like you found your favorite toy, and you lost it when you were a little kid, and next thing you know a year later you actually found it, and you got this feeling, you know?

Estevan's love for science stretches back to his childhood, but it became dormant during his early schooling experience. He cited Ms. K's class as rekindling his love of science, thanks in part to her requiring him to figure answers out for himself. This rekindling makes it more likely for Estevan to stick with science. And there are hints that he is now projecting science into an important role in his future as well, saying that "it's gonna stay with me," and that he is "not gonna give up on science". This is promising, given research that finds students about this age who think about science as figuring into their future are more likely to pursue a degree or a career in science (Tai et al., 2006). This rekindling of his lost feeling towards science is powerful for Estevan. But what happened in Ms. K's class, specifically, that mediated this rekindling for Estevan? This is the question we turn to in Part 3.

**Part 2 – Summary & Discussion**

   Through interviews with Estevan and Ms. K., we found that his persistence was explained in part by an alignment between his sense of self as someone who loves challenges, and his epistemological stance that science involves the challenge of figuring out stuff.  We also learned that Estevan's interest and persistence in science may extend to longer timescales, as he expressed aspirations to pursue a career in science and was already formulating concrete plans to make that happen.  He also described early interactions with his father through which he grew to love science and see it as involving figuring things out for yourself.  This feeling and epistemological stance toward science was diminished in school until he reached Ms. K's class.

   In Part 3, we aim to understand more about the learning environment and the dynamics in Ms. K's classroom that supported Estevan's rekindled interest in science.  What about Ms. K's class, and Estevan's interactions with Ms. K and others, sparked his engagement?

**DATA & DISCUSSION – PART 3**

   In Part 2, we analyzed interviews with Estevan and Ms. K to show how Estevan's sense of self as a lover of challenges intersected with his epistemological stance toward science as involving the challenge of figuring things out.  In Part 3, we examine whether and how this intersection played out in the classroom.  First, we demonstrate that Estevan's sense of self as someone who loves a challenge is corroborated by his enacted identity in the classroom, in how he responds to challenges and poses challenges of a social and intellectual nature to others.  We also show how Ms. K supports Estevan (and the other students) in coming to see science as involving the challenge of figuring things out for yourself.  In this way, we see how Ms. K creates an environment that supports Estevan's persistent engagement.  In her class, his challenging attitude is seen as not a barrier to doing science, but as a central part of being a scientist.  We illustrate how she does so both through the mutually playful and caring relationship both Ms. K and Estevan establish, but also through creating a classroom atmosphere in which feistiness is welcome and students are free to debate their differing mechanistic ideas.

**Estevan playfully challenged Ms. K in class**

   In addition to rising to challenges posed to him, as seen previously, Estevan's professed love of challenges also played out in the classroom as he directed playful challenges at Ms. K.  These challenges did not always involve disciplinary content and practices. For example, in one class session early in the year, Ms. K was transitioning from one activity to another and counting down the time until students needed to be back in their seats.  Estevan started inserting random numbers into her countdown, and Ms. K started skipping around in her counting too, from eighteen back up to twenty-five.  He gave her a rough time about it: "Twenty-five again?  You're skipping a lot of numbers. You can't count."  Ms. K could have taken this as a disrespectful dig, but instead she kept counting inaccurately, suggesting that she likely took up Estevan's commentary as a playful tease.

In another example from early in the school year, Ms. K was reviewing the periodic table with her class before a test.  She posed the review activity as a challenge: she challenged them to tell her everything they could about oxygen, just based on its entry in the periodic table.  When Ms. K called on Estevan, he playfully challenged her back, pointing to oxygen's entry on the periodic table throwing the question back at her: "I have a question; What can you tell me about that (oxygen)?" Ms. K fired back, "everything." Estevan continued to challenge to provide examples, but Ms. K retorted "After you tell me I'ma tell you." Ms. K would not relinquish on her challenge to have the students come up with their own answers.

**Estevan Challenged Other Students**

Estevan's enacted identity as a lover of challenges seems to involve challenging other students in intellectual debates, a behavior Ms. K supported. In these debates, Estevan oriented to both the challenge of figuring things out during classroom discussions and the challenge of winning an argument against his classmates. These were mutually reinforcing at times, but sometimes Estevan's focus on winning trumped constructively working to figure things out. Here, we provide a few examples from an argument between Estevan and another student (Reggie) during the seasons discussion, illustrating these dual foci for Estevan and how Ms. K supported him in finding a productive balance.

Ms. K had posed the question of how the Earth could be closer to the Sun in the winter. Reggie proposed an idea about the Moon making the Earth colder in winter. Ms. K asked him to clarify, "How does the Moon have to do with the heat of the Earth?". Reggie clarifies (line 23). Ms. K presses him to clarify further before asking who agreed with Reggie (line 26), and Estevan indicated that he did not (line 27):

> 23. Reggie: At night, so maybe in the day, it's kinda opposite. The- the-the Moon affects the Sun so it kinda makes it kinda colder, if the Moon is like blocking the sunlight.
> 24. Ms. K: Oh, I thought you said reflects it. If it blocks it- Reggie! If the Moon blocks the Sun, wouldn't it be dark then?
> 25. Estevan: That's an eclipse.
> 26. Ms. K: ((to Reggie)) Okay, that's your idea? ((to class)) Who agrees with Reggie?
> 27. Estevan: I don't.
> 28. Ms. K: You don't?
> 29. Estevan: No.
> 30. Ms. K: What's your idea, Estevan?
> 31. Estevan: I agree with her ((another student))... because it's true what she said though, I mean the Earth is kinda tilted.

Reggie then challenged Estevan back by conceding that while the Earth is tilted, the tilt does not explain the seasons. In the exchange that follows, Estevan and Reggie got into a back-and-forth so feisty that other students tried to calm them down:

> 32. Reggie: Can I say something, Ms. K?
> 33. Ms. K: Okay Reggie, what do you have to say?
> 34. Reggie: Even though the Earth is tilted, that has nothing to do-
> 35. Estevan: Yes it does.
> 36. Reggie: That has nothing to do with it. Like, I think, that like right now, during the day, you know, China is night right now. But yeah, so- yeah, so, just- but that kind of proves my point. You know, hotter, because-

37. Estevan: Um, um, I'm about-
38. Lainie: You just proved yourself wrong.
39. Estevan: Okay I'm going to tell you something ((chuckles)), you did just prove yourself wrong... first of all the Earth's axis is tilted, that means that part of the Earth is actually closer to the Sun.
40. Reggie: Okay but the Earth rotates ((makes spinning gesture))-
41. Estevan: ((talking over Reggie)) Then the other part is actually far away, which would mean the part that's closer will be hot, right? Right?!
42. Reggie: It'd be colder.
43. Tanisha: Why you gotta be so loud? Calm down.

Tanisha's comment in line 43 ("Why you gotta be so loud?") suggests that she (and perhaps other students) viewed Estevan's feistiness as excessive. Not only did he get loud, but he also got so insistent with his questioning that he did not let Reggie finish his turn of conversation.

Rather than reprimanding Estevan, Ms. K supported his feistiness: "Sometimes we get a little louder when we're trying to explain. You guys know how I get, you wanna prove your point, okay?" Through this, she communicated that a certain amount of feistiness was to be expected during passionate conversation. However, she subsequently refocused students on the substance of the argument between Estevan and Reggie, and she shortly thereafter halted the argument when it became so overlapping in nature that the students were no longer hearing each other. In other words, Ms. K was supportive of Estevan and others' challenging and feistiness while the conversation was productive, but when it got to the point that they were not listening to each other's ideas, she pulled them back. This correction on Ms. K's part did not shut down Estevan's or the other students' sensemaking, in fact their substantive debate continued on for 20 more minutes. Along the way Ms. K had to stop the debate several more times to remind them of the norms of conversation (not talking over each other).

This feistiness corroborates with what Estevan said in his interview, when he how he gets really "serious" with group discussions:

I mean, we talk, like she makes us do the little debate, like you know, talk about- you saw that last time, how we were talking about. Actually, for me, like, getting the discussion with the classroom, I get pretty serious with it. ((Estevan & Jen laugh)) I get pretty serious with it. I mean, like, literally serious.

This seriousness is in contrast with his playful exchanges with Ms. K. Moreover, the difference is at least in part epistemological—he gets "serious" when he thinks someone else is arguing even though they know they're wrong:

Because sometimes, then I- if I'm trying to say that they're wrong and that I'm right, but they don't want to actually admit it. So I'll be like you're just saying the stuff that I'm saying right now, so you're actually admitting that you're wrong, and you know, let's try to go the other way. I'm like, I get really serious with it because, you know, if I know somebody's actually wrong, I will actually correct them and stuff. And if I know that they know

```
something that I don't, I mean, I'm like, okay I've learned something new,
how do you know that?  Like, how do you find out that?  Where do you find
that at so I can learn about it too?
```

This description of his frustration and its source was likely behind his intense
debate with Reggie.  For instance, in line 39, Estevan communicated that Reggie had just
proved himself wrong.  From that point on, Estevan may have been operating under the
assumption that Reggie *knew* he was wrong and was arguing unnecessarily.  He says so
directly to Reggie later on in the debate.   Thus, although Estevan usually responded
substantively to others' ideas during debates, evaluating them in terms of whether they
make sense or not, this frustration kicked in and got in the way of his making sense of (or
even listening to) what Reggie said.

Ms. K's work on establishing the norms of their class discussions to allow for
feistiness provided an opportunity for Estevan to utilize his love of challenging people in
discussions that focused on making sense of scientific phenomena.  In letting him go
when he was feisty, but stopping him when he stopped listening, he may have gotten the
message that this is the sort of discussion that is not only tolerated but valued, as long as
people are able to be heard.

**Ms. K Supports the Connection Between Estevan's Identity & Epistemology**

*"You ARE the scientist"*
The whole class discussed the seasons for 30 minutes before Ms. K starts
transitioning them to the next phase of the activity, modeling the Sun and Earth to prove
their explanation of how it can be that the Earth is closer to the Sun in the winter.

Over the course of the discussion thus far, they have touched upon various aspects
of the seasons, driven largely by the debate between whether rotation or tilt is what
causes the seasons.  The debate has been quite contentious at times, especially between
Estevan and Reggie.  Ms. K has to stop the discussion at several points throughout the
half hour to remind everyone not to talk over each other.

Ultimately, Ms. K utilizes their contentious debate as motivation for the next
activity.  After all this back and forth, instead of giving them the answer, she challenges
them to think about how they can find out.

```
                ((students debating, talking  over each other))
        Ms. K:  Okay. Alright! ((students still talking over each other)) STOP.
                STOP.  ((students turn to Ms. K)) –HOW– are we gonna find
                this out.
      Student:  Internet.
     Students:  ((almost in unison)) Google!
        Ms. K:  I don't- I don't –want– Google.  I want-
                ((Ms. K waits for silence.  Students settle down.))
        Ms. K:  –How– could we find out the answer to this.  What could you
                do?  And do –not– say look it up on the internet, because it's
                not available.
      Student:  Make a diagram.
        Ms. K:  Make a diagram…
```

```
Estevan:     Just call a scientist.
Courtney:    Yeah, just go to a scientist.
Estevan:     Go to a scientist.
Ms. K:       No.
Emanuel:     Watch the sky!  ((Students chuckle))
Ms. K:       I mean an **investigation**, how could you d-
             ((Students start talking, chuckling at Emanuel's answer))
             ((Ms. K pauses, waits for them to stop talking))
```

It seems to have become a different kind of conversation here. Up until this point, in the seasons discussion she did not judge the students' ideas about what could cause the seasons, only tried to clarify them.  But here, the question has become "how do we find out?"  Ms. K sends the message that for this question, some answers are better than others.  They wanted to look up the "internet", "Google", and "ask a scientist", but she did not want them to defer to an authority.   When it was clear they were not picking up on what she had in mind, she told them she wanted them to investigate for themselves.

The students get distracted before Ms. K can finish her thought, so she pauses and waits to make sure everyone is listening when she tells them what they will be doing next:

```
             ((As Ms. K waits, the students quiet back down.  Ms. K then
             waits for several seconds *after* students are silent))
Ms. K:       We're not going to talk to scientists, you –**are**– the scientists.
             **You**, in your group are gonna figure out –how– ... Let me tell
             you what you're gonna do, you're gonna figure this out.  You're
             gonna demonstrate it to the classroom. You are NOT gonna go
             to Google and look it up.
Estevan:     You got the little thingy ball like the Earth?
Ms. K:       I've got it!  I've got it, got it –for– you.
```

By telling them "You –**are**– the scientists. **You** in your group are gonna figure this out," Ms. K draws a connection between *figuring it out for yourself* and *being a scientist*. She tells them that they will figure out and demonstrate to the classroom.  Estevan seems eager to get started, asking if she has something ball-shaped to model the Earth.  In framing the activity as a challenge to 'be the scientist' by 'figuring it out', Ms. K is setting up an environment that supports Estevan's persistent engagement by supporting his identity as a lover of challenges, and supporting his epistemological stance that science involves the challenge of figuring things out.

### *"You Gotta Give Me A Clue Or Something"*

Even as Ms. K is transitioning them to the modeling activity, the students keep bringing up new ideas and questions pertaining to the seasons discussion.  She hears each idea out, but gently steers the conversation back to the modeling activity.  Estevan asks a question about why deserts are so hot in the day and cold at night, but when another student challenges him, he snaps back.  Instead of reprimanding his feistiness, she told everyone *what he was saying* was right, but the *why* is missing and she's not going to tell:

44. Ms. K: Okay, I'm gonna give you a globe, an inflatable globe, yes.
45. Estevan: Yeah I got a question.
46. Ms. K: ((to another student)) If you can model it-
47. Estevan: Ms. K, I got a question.
48. Ms. K: Shh.
49. Estevan: Nah, I just have a question.
50. Ms. K: Shh.
51. Estevan: Tell me why, um-
52. Ms. K: ((turns to Estevan)) Hmm?
53. Estevan: I got a question.
54. Ms. K: I wanna hear it.
55. Estevan: Um, I said um, tell me why, you know, Mexico right, the desert over there, tell me why in the day it's like really hot, but at night it's like freezing.

Estevan's question represented a challenge both to moving the activity forward and to the dynamic of figuring things out, as he implored her to "tell me why" repeatedly (lines 51 and 55). Ms. K responded by acknowledging Estevan's question but refusing to answer it: "What's he's saying is actually true. Why, I'm not gonna tell you. Since this is a different conversation, we are gonna stop right now." She then returned to the activity she was trying to get started:

56. Ms. K: Here's what I'm gonna do, I'm going to give each group a globe, you're gonna use that globe, and any other models in the classroom that you want to use. You have to explain why part of the- uh, you have to explain why the the Earth is closer in the winter, to the Sun. You have to prove it. And explain it.
57. Estevan: Aight so first you gotta give us a clue or something.
58. Ms. K: I'm not giving you anything, you have your brain ((taps Estevan on the shoulder)).

Estevan conceded the shift in activity but still pushed back on the epistemological dynamics by insisting she give them a clue (line 57). Ms. K responded by indicating that Estevan has the necessary resources to answer the question (i.e., his brain) (line 58), ultimately maintaining her epistemological line.

## Part 3 – Summary & Discussion

In the various episodes, we find that Estevan's sense of self as someone who loves a challenge is corroborated by his enacted identity in the classroom. He not only responds to challenges by digging in and persisting through them, he also loves to challenge others, including Ms. K. For her part, Ms. K also likes to challenge others. When they set this challenging attitude toward each other it establishes a mutual caring relationship that incorporates this playful challenging. Ms. K also sets up a classroom environment in which feistiness is valued, and in which students are encouraged to figure things out for themselves. She connects figuring things out with what it means to do science. This environment supports Estevan in connecting up his identity as a lover of challenges with a personal epistemological stance that science involves the challenge of figuring things out. This helps explain his persistent engagement in science.

Estevan's tendency to challenge both his classmates and especially his teacher might be interpreted by many teachers as a sign of disrespect. For instance, Estevan mocked Ms. K's counting ability when she skipped numbers as she counted down to the next activity. When Ms. K tasked students with telling her everything they know about oxygen based on its periodic table entry, Estevan flipped the question on its head by challenging *her* to tell *them* everything she knows about oxygen, knowingly undermining the activity. In class discussions, Estevan got so feisty that other students made moves to repair the conversational norms (Tanisha in line 43, "Why you gotta be so loud?"). Any one of these manifestations of Estevan's challenging behaviors would push against the grain of what many consider to be acceptable classroom behavior. It would be easy to imagine another teacher viewing such moves as reprimandible, or even taking the sum total of these moves to label Estevan as a "problem student." In fact, at the beginning of the year, Estevan had such a reputation.

Yet Ms. K did not discipline Estevan for these behaviors. When Estevan mocked her counting, Ms. K let it slide, and even played along. When Estevan challenged her by asking what she knows about oxygen, she countered the challenge by stating that she knows "everything" and playfully turned the challenge back on him ("After you tell me, I'ma tell you"). When Estevan got feisty in discussion, Ms. K actually came to his defense, acknowledging that he was "trying to explain" and even identifying with his feistiness ("You know how I get"). This afforded Estevan the space to tap into his love of challenging himself and others, which he was able to recruit (with Ms. K's support) toward working to understand scientific phenomena. Thus, Ms. K's classroom environment allowed Estevan to express himself in ways that positively reinforced his sense of self as a lover of challenges and marshaled his enactment of this identity toward productive scientific ends.

## Part 3 – Epilogue

Another example of Estevan striving to make sense of a scientific scenario occurred later in the year, in May 2012. Jen (second author) was a guest instructor in his science class for the day, and she engaged students in thinking about a conceptual physics question modeled off the "Thalia on a Swing" problem from *The Physics Teacher* (Hewitt, 2009). The problem involves a girl named Thalia swinging on a swing, reaching a certain height ($H_1$) with each swing. One time, as she swings forward, she picks up her backpack from the ground and continues swinging – is the height she reaches after picking up the backpack ($H_2$) lower than, equal to, or higher than $H_1$, and why? Then, as she swings backward, she drops the backpack to the ground and continues swinging – how does the height she reaches after dropping the backpack ($H_3$) compare to $H_1$ and $H_2$, and why?

As Estevan's group (himself and two other boys) started to grapple with the first part of the question (Thalia picking up the backpack and the relationship between $H_1$ and $H_2$), they disputed the initial assumption that $H_1$ would be the same for forward and backward swings. As Estevan noted, when you're going forward, "your whole body's into it. But when you're going back, your whole body is not into it because you're just coming back, so you're very light." He seemed to be building from an experience or intuition that your whole body puts in effort to go forward on a swing, but when you go back, it's "just your legs, you just pull them back again." This led Estevan and his group to believe that Thalia's swing going forward would be higher than her swing going backward, so there was no consistent $H_1$ – in essence, challenging the very set-up of the problem.

As discussion continued, the class generally agreed that $H_2$ would be lower than $H_1$ because the backpack would weigh Thalia down and/or mess up her motion as she reached down to pick it up, preventing her from swinging as high as she usually would. There was less consensus on how $H_3$ would compare, though – students thought Thalia might swing higher again because the extra weight would speed her up on the way down, and the backpack flying off would propel her body upward. Yet Estevan remained concerned about the backpack's impact on the smoothness of Thalia's swinging motion, agreeing that she might go higher again but not as high as other students thought because "it would mess up your motion when you're going back, you know, cause you gotta drop the bookbag."

In this example, Estevan was focused on making sense of the motion of the swing under different circumstances. He demonstrated a willingness to voice and go with his own intuitions in the face of other students' ideas and the set-up of the problem itself (and, later, me) – not deferring to potential authority figures. This example, which came later on in the year, demonstrates that Estevan's persistent engagement with the practices of science extended beyond just the seasons discussion. The fact that he engaged deeply in sensemaking in this context illustrates that his persistent engagement was not due simply to an interest in a particular topic (seasons), nor dependent on his access to Ms. K (although it is of course possible he was trying to impress Jen).

## SUMMARY & DISCUSSION

In part I, we recounted the classroom episode that brought Estevan to our attention, because of his persistent engagement in understanding a phenomenon despite his obvious frustration. In part II, we found an explanation of this persistence—that Estevan sees himself as someone who rises to challenges, and that he sees science as involving the challenge of figuring things out for himself. In part III, we saw how this dynamic played out in the classroom (some of which we had videotaped before Estevan jumped out to us as an exemplary case of someone suddenly (re)hooked into science). Over the course of a year, there is evidence that Ms. K supported Estevan in (re)connecting this view of science as a challenge of coming up with explanations of mysterious phenomena and with his identity as someone who loves a challenge, and who conversely loves to challenge others.

Previous research has suggested that one way to reach kids otherwise turned away from science was through a topical connection. But Estevan's enthusiasm does not seem to be particularly connected to the topic so much as the challenge of figuring things out, and the challenge of making sense of other peoples' ideas, putting his own ideas out there, and defending them in the ensuing discussion.

# IMPLICATIONS & FUTURE DIRECTIONS

## Implications for Educational Research

Our case study of Estevan adds to the literature on engaging underrepresented students in science. Previous work has emphasized connections between aspects of students' identities and specific scientific content or culturally-relevant epistemologies/ways of knowing (Bang & Medin, 2010; Hudicourt-Barnes, 2003; Warren et al., 2001). Here, we provide another way in which epistemology can serve as a link between interest in science and identity. In his interview, Estevan communicated a sense of self as someone who loves challenges, which may or may not tie to his cultural background as an immigrant from Central America. During the focal episode, Estevan perceived as challenges 1) making sense of the tilt of the Earth in the fall and spring, and 2) grappling with Ms. K's question about why the Sun is closer to us in the winter than in the summer. His sense of self as someone who faces challenges head-on was enacted and coupled, at least in these instances, with a personal epistemological stance that science involves the challenge of figuring things out for yourself.

An alternative explanation is that Estevan has a particular interest in topics related to the Earth and/or the movement of celestial bodies, including how we have seasons, and that this topic-specific interest explains his enjoyment and persistence. However, during the course of the interview, Estevan also raised questions about chemicals, plants, rocks, etc. In class, he grappled with chemistry topics (e.g., how exothermic reactions occur) and physics topics (e.g., explaining the motion of a girl on a swing) with equal enthusiasm. What is common across these various topics, that made them engaging for Estevan, is that they provided him with challenging opportunities to figure out complex problems.

It is important to note that we are not arguing that this epistemological stance of figuring things out for yourself is universally productive. In fact, we see part of its productivity for Estevan as living *in* its alignment with his love of challenges. Yet this same alignment might be counterproductive for students who shy away from challenges. Thus, while we highlight this particular intersection between identity and epistemology as generally productive for Estevan's engagement in science, we call for more in-depth case studies of activated science learners (underrepresented and otherwise) to enhance our understanding of the underlying dynamics. Such case studies, drawing on data from numerous contexts, would allow for the recognition of individual patterns and eventually facilitate comparison across cases for commonalities and distinctions.

## Implications for Science Teaching

A critical point to acknowledge is that Ms. K's classroom environment facilitated Estevan's scientific engagement in a way that his previous science classes had not. We suggest that an important feature of Ms. K's teaching was her close attention and openness to students' ideas and approaches in class.

For instance, although Ms. K pushed back on students' ideas, she was attentive to them and open to considering ideas that were not canonically correct. When several students raised the idea that the Moon was closer to the Earth in the winter, making it colder, Ms. K authentically engaged with their idea. She did not correct it, but rather sought to understand their perspective, letting them know when she did not understand and asking probing questions like, "How does the Moon have to do with the heat of the Earth?". Moreover, when she understood that the students were thinking of the Moon blocking the Sun, her challenge directly addressed this point: "How do you explain that for a sunny day?" Even in her pushbacks, her focus was apparently not on whether the idea was correct or not, but whether it made sense (at least the students took it up, since they kept elaborating on their ideas). These were the standards to which she held students, tacitly communicating that these were the standards by which they should assess their own ideas in science (Coffey, Hammer, Levin, & Grant, 2011). Focusing students on the sensibility of their own ideas instead of the "right answer" was not an easy transition, as evidenced by some of Estevan's challenging – yet this is ultimately what Estevan came to see as empowering and "fun."

Additionally, we previously acknowledged that some of Estevan's challenging behavior could have been perceived as problematic. Rather than disciplining Estevan, though, Ms. K worked *with* Estevan to channel his challenging into productive scientific ends. In other words, she was open to his approach and to a core aspect of *him*, and she worked to connect what he brought to the table with science. This kind of openness to the multiplicity of connections that might exist between students and science is of critical importance in the science classroom if we wish to encourage students' continued interest and engagement in science.

**Future Directions for Research**

Better understanding the dynamics between Estevan and Ms. K raises other questions for us. First, how stable is Estevan's interest in science? As he moves into high school, he will likely face science classes more like the classes he experienced prior to Ms. K's – classes in which the focus is on the right answer and what the textbook says rather than students' ideas and ingenuity. How will he respond if he takes classes in which his love of challenges is not only less relevant, but possibly disciplined? We will attempt to follow up with Estevan to get a sense of how he is experiencing science in high school. Moreover, this raises a more general salient issue – the need to create mechanisms for following up with students on a longitudinal timescale to better understand the impact of given classes.

Second, how did Ms. K come to open up this kind of space in her classroom? Earlier, we alluded to Ms. K's own frustration when engaging in an inquiry on circuits as part of her participation in the professional development project. In a subsequent paper, we will examine video of her participation and interaction with the facilitator during this inquiry, as well as her reflections on how this experience shaped her classroom practice.

**Future Directions for Estevan**

As Estevan moves on to high school, he may not end up with science teachers who he feels value his approach to science. There is a chance they might interpret his playful challenging as a sign of disrespect. There is also a chance the teachers he has in high school will not support his epistemological stance of finding things out for yourself. As he said, his previous teachers had not let him do what he wanted to do, find out. But once he got a teacher who allowed space for him to figure things out for himself, and who posed it as a challenge for him—the challenge of doing science—he finds his fire rekindled.

Hopefully, now that he is older and perhaps more self-aware of this love of science, he can keep the fire going it is rekindled for good.  In his favor, he has no intentions of letting his re-found love of science go away anytime soon:

> Because I mean science, for me, it'll stay with me until I die.  Like I said, science for me, it's gonna stay with me, I mean, I'm not gonna give up on science though.  They already took that feeling away when I was in elementary school, and she [Ms. K] brought it back in though, I mean, like something that you found, like you found your favorite toy, and you lost it when you were a little kid, and next thing you know a year later you actually found it, and you got this feeling, you know?

He will face many obstacles along the way in his pursuit of a degree in science and eventually a career, but if he is able to stably frame these obstacles as a series of challenges to overcome he may sustain his interest.

## ACKNOWLEDGEMENTS

The authors would like to thank Jessica Watkins, David Hammer, and Megan Luce for helpful comments and suggestions.  This work was supported by NSF MSP 0831970, as well as the Gordon and Betty Moore Foundation.